\documentclass[showpacs,amsmath,amssymb,twocolumn,nofootinbib]{revtex4}
\newcommand{\be}{\begin{equation}}
\newcommand{\ee}{\end{equation}}
\newcommand{\Hds}{\mathcal{H}}
\newcommand{\bk}{{\bf k}}
\newcommand{\bx}{{\bf x}}
\newcommand{\du}[3]{\mbox{$#1^{\,\;#2}_{#3}$}}
\begin{document}
\title{Vacuum energy and spectral function sum rules}
\author{Alexander Y. Kamenshchik}
\affiliation{Dipartimento di Fisica and INFN, Via 
Irnerio 46, 40126 Bologna,Italy\\
L.D. Landau Institute for Theoretical Physics of the 
Russian Academy of Sciences, Kosygin str. 2, 
119334 Moscow, Russia}
\author{Alessandro Tronconi}
\affiliation{Dipartimento di Fisica and INFN, Via 
Irnerio 46, 40126 Bologna,Italy}
\author{Gian Paolo Vacca}
\affiliation{Dipartimento di Fisica and INFN, Via 
Irnerio 46, 40126 Bologna,Italy}
\author{Giovanni Venturi }
\affiliation{Dipartimento di Fisica and INFN, Via 
Irnerio 46, 40126 Bologna,Italy}
%
%\date{1 Dic. 2006}
%

\begin{abstract}
We reformulate the problem of the cancellation of the ultraviolet divergencies of
the vacuum energy, particularly important at the cosmological level,
in terms of a saturation of spectral function sum rules which leads to a set
of conditions on the spectrum of the fundamental theory.
We specialize the approach to both Minkowski and de Sitter space-times and
investigate some examples. 
\end{abstract}

\pacs{98.80.Cq,95.36.+x,04.62.+v,11.55.Hx}

\maketitle 
\section{Introduction}
The energy density of the vacuum, and in general the cosmological constant problem,
has always attracted much attention \cite{Zeldovich:1967gd,Weinberg:1988cp}. 
The recent discovery of cosmic acceleration \cite{accel} has stimulated
the discussion about what kind of dark energy could be responsible for this
phenomenon \cite{dark}.
The cosmological constant is the most natural candidate. The enigma arises
because the zero point energy associated with quantum fields and naively
calculated by introducing a cutoff at the Planck scale appears to be some 118
orders of magnitude bigger than the observable value of the 
cosmological constant energy density. Many explanations have been put forward
for this \cite{Nobbenhuis:2004wn}, prominent among them are symmetry mechanisms and in 
particular supersymmetry. Indeed since the contributions to the ground state energy of 
Fermions and Bosons have opposite signs unbroken supersymmetry leads to a
vanishing vacuum energy. Unfortunately supersymmetry, if it exists, certainly
must be broken and at this point one appears to still be left with the
cosmological constant problem. One lesson one learns however is that the number of fermionic
and bosonic degrees of freedom must be equal, if one wants to have an exact cancellation at 
least of the quartically ultraviolet divergent part of the vacuum energy\footnote{The finite 
part of the vacuum energy (which is difficult to calculate explicitly) could be responsible 
for the small observable value of the cosmological constant}.\\
%%%%%%%%%%%%%%%%%%%%
In order to employ what one has learned from the above one may seek
inspiration from the study of other symmetry breaking, in particular chiral symmetry.
Indeed if chiral $SU(2)\times SU(2)$ was an exact symmetry one would have
expected the $\rho$ meson to be accompanied by an axial vector meson of the
same mass, which certainly is not the case. A way out of this was found
through the introduction of an asymptotic chiral symmetry
\cite{Weinberg:1967kj} leading to spectral function sum rules which are
related to the short-distance behaviour of products of vector and axial-vector
currents \cite{Bernard:1975cd}. The approximate saturation of the spectral
functions with suitable combinations of particles lead to a physically
satisfactory relation between vector and axial-vector meson masses.\\
%%%%%%%%%%%%%%%%%%%%%
We wish to follow an analogous procedure for the case of vacuum energies.
Indeed we shall first obtain expressions for the large momentum behaviour of
the vacuum energy in de Sitter space for arbitrary mass, spin 1/2 Fermions and
spin 0 and spin 1 Bosons. This is done in the next Section. Subsequently we shall,
through the use of spectral functions, represent the contributions of an
arbitrary number of Fermions and Bosons. On examining the contributions of all
particle species to the vacuum energy and requiring that all divergent
contributions cancel, we obtain constraints between diverse spectral functions
and their moments. This of course leads to contraints on the mass spectra of
the particles which approximately saturate the sum rules. These constraints
are studied in Section III. Lastly our results are summarized and discussed in the Conclusions.

%%%%%%%%%%%%%%%%%%%%%%%%%%%%%%%%%%%%%%%%%%%%%
\section{Vacuum energy of free fields}
In what follows we shall consider a flat Friedmann universe with the Robertson-Walker metric 
\be
ds^2 = -dt^2 +a^2 (t) dx^i dx^i, 
\label{rwmetric}
\ee
where the cosmological radius $a(t)$ obeys the de Sitter expansion law:
$a(t)=a_0{\rm e^{\Hds t}}$. We shall calculate the ultraviolet divergent
contributions to vacuum energy for free fields on this spatially flat de Sitter background.

\subsection{Bosonic fields}
Free bosonic field dynamics, on a Friedmann-Robertson-Walker (FRW) space-time,
is known to be equivalent
\cite{birrell} to that of an infinite system of time dependent harmonic
oscillators (TDHO). Subsequently, at the quantum level, one can express the
vacuum expectation value of the field Hamiltonian
in terms of the so-called Ermakov-Pinney variables \cite{lewis,Ermakov,gao}. These
variables are introduced to define the Invariant operators one uses to build
the Hilbert space of the solutions for the TDHO. Since we should be interested
in the ultraviolet divergent contributions to the Bunch-Davies vacuum energy
\cite{bunch}, we shall finally consider the modes up to an U.V. cutoff, and,
in particular, restrict the analysis to trans-Planckian
$k$-modes of the fields, namely the limit $z\equiv\Hds^{-1}k/a\gg1$.

%%%%%%%%%%%%%%%%%%%%%%%%%%%%%%%%%%%%%%%%%%%%%
\subsubsection{Method of Invariants}
When the action of a quantum system can be written in the form of some TDHO\be\label{STDHO}
S=\int dt\frac{1}{2}\left(\frac{\dot q^2}{F(t)}-G(t)\,q^2\right)
\ee
with Hamiltonian given by
\be\label{HTDHO}
\hat H=\frac{F}{2}\hat p^2+\frac{G}{2}\hat q^2,
\ee
where $\hat p=\dot{\hat q}/F$,
 $F$ may be interpreted as an inverse effective mass of the harmonic oscillator while 
$G$ is the mass multiplied by the effective frequency squared. 
The solutions to the Schr\"odinger equation are eigenstates of the quadratic
invariant operator $\hat I$ defined by
\be\label{invariantdef}
\frac{\partial \hat I}{\partial t}=i\,\left[\hat I,\hat H\right].
\ee
The quadratic invariant operator for the case (\ref{HTDHO}) is given by
\be\label{invariant}
\hat I=\frac{1}{2}\left[\frac{\hat q^2}{\rho^2}+\left(\rho\hat p-
\frac{\dot\rho\,\hat q}{F}\right)^2\right]
\ee
where $\rho=\sqrt{F}x$, $x$ is the Ermakov-Pinney variable satisfying the equation \cite{Ermakov}
\be\label{Pinneyeq}
\ddot x+\Omega^2x=\frac{1}{x^3}
\ee
and $\Omega^2=F G+\frac{\ddot F}{2F}-\frac{3\dot F^2}{4F^2}$.
Both the invariant (\ref{invariant}) and the Hamiltonian (\ref{HTDHO}) can be
written in terms of time dependent creation-annihilation operators but, in the
time dependent case, the Hilbert space of the solutions of the Schr\"odinger
equation is generated by the linear invariant creation-annihilation operators;
energy eigenstates are related to these solutions by a Bogoliubov rotation
depending on time through $\rho$, $F$ and $G$.
The invariant vacuum $|0_I\rangle$ and the invariant operator
(\ref{invariant}) itself are not unique but depend on the initial conditions
chosen for the Ermakov-Pinney equation; in any case the expectation value of
the Hamiltonian (\ref{HTDHO}) with respect to it can be expressed as
\be\label{vevH}
\langle 0_I|\hat H|0_I\rangle=\frac{1}{4}\left[\frac{\dot \rho^2}{F}+G\rho^2+
\frac{F}{\rho^2}\right]
\ee
and is a function of time. Setting $\rho^2(t_0)=\sqrt{F(t_0)/G(t_0)}$ and
$\dot \rho(t_0)=0$, the invariant vacuum $|0_I\rangle$ 
(which is related to the Bunch-Davies vacuum)
coincides with the Hamiltonian vacuum $|0\rangle$ at time $t_0$.\\
Technically the full quantum evolution is determined once the Ermakov-Pinney
equation is solved.
The exact solution of the Ermakov-Pinney equation is known if the homogeneous,
linear equation associated with it is solvable; otherwise the Ermakov-Pinney
equation itself can be solved perturbatively and gives the correct adiabatic series.\\
In particular, when handling free field evolution in de Sitter space-time\cite{bertoni},
the adiabatic parameter should be chosen to be the $z$ defined above. In the
ultraviolet regime, one must set $x=\xi/\sqrt{z}$, where $\xi$ satisfies
\be\label{Pinneypert}
\frac{d^2\xi}{dz^2}+\frac{1}{z^2}\left[\Hds^{-2}\Omega^2\left(z\right)+
\frac{1}{4}\right]\xi=\frac{1}{\Hds^2 \xi^3},
\ee
and the modified Ermakov-Pinney equation (\ref{Pinneypert})
gives the correct adiabatic solution.
%%%%%%%%%%%%%%%%%%%%%%%%%%%%%%%%%%%%%%%%%%
\subsubsection{Real scalar field}
The action for a minimally coupled, massive scalar field $\Phi(\bx,t)$ is
\be\label{Sscalar}
S [\Phi]=\int d^4x\sqrt{-g}
\left[-\frac{1}{2} g^{\mu \nu} \partial_{\mu} \Phi \partial_{\nu} \Phi 
- \frac{1}{2} m^2 \Phi ^2\right];
\ee
if one considers the Fourier transform of the field
\be\label{fourscalar}
\Phi(\bx,t)\equiv\frac{1}{\sqrt{2V}}
\sum_{\bk}{\rm e}^{i\bk \cdot \bx}\left[\phi_1(\bk,t)+i\phi_2(\bk,t)\right]
\ee
where $\int d\bx=V$ and $\phi(\bk,t)$ is a real function, the action
(\ref{Sscalar}) can be rewritten as
\be\label{Skscalar}
S=\frac{1}{2}\sum_{i=1,2}\sum_{\bk}
\int \frac{a^3}{2}dt\left(\dot\phi_i(\bk)^2-\omega_k^2\phi_i(\bk)^2\right)
\ee
where $\omega_k^2\equiv \frac{\bk^2}{a^2}+m^2$. We note that one can eliminate
the overall factor $1/2$ in (\ref{Skscalar}) by observing that
$\phi_1(\bk)=\phi_1(-\bk)$ and $\phi_2(\bk)=-\phi_2(-\bk)$  thus restricting
the sum over the subspace $\bk^+$ of the independent degrees of freedom. 
For each independent mode $\bk,i$, one then obtains the following Hamiltonian
\be\label{hamkscalar} 
\hat H_i(\bk)=\frac{\hat \pi_i(\bk)^2}{2a^3}+\frac{a^3\omega_k^2\,\hat \phi_i(\bk)^2}{2}
\ee
where $\hat \pi_i(\bk)\equiv a^3\dot{\hat\phi}_i(\bk)$; the full Hamiltonian
can then be written as $\hat H_S=\sum_{i=1,2}\sum_{\bk^+}\hat
H_i(\bk)$. Finally, to calculate the vacuum expectation value (\ref{vevH}), in
the continuum limit
\be\label{contlim}
\frac{1}{V}\sum_\bk\stackrel{V\rightarrow\infty}{\longrightarrow}
\frac{1}{(2\pi)^{3}}\int k^2dk\,d\Omega,
\ee
it is sufficient to just keep  $\hat\phi\equiv\hat \phi_1$ and integrate over
the complete solid angle
\begin{eqnarray}
\label{Hcontscalar}
&&\!\!\!\!\!\!\!\!\!\!\!\!\frac{\langle 0_I|\hat H_S|0_I\rangle}{V} 
\stackrel{V\rightarrow\infty}{\longrightarrow}
\frac{1}{2\pi^2}\int\limits_0^\infty k^2E_0^S(k)dk\nonumber \\
&&\!\!\!\!\!\!\!\!\!\!\!\!=\frac{1}{2\pi^2}\int\limits_0^\infty k^2
\langle 0_I|\left(\frac{\hat \pi(\bk)^2}{2a^3}+\frac{a^3\omega_k^2\,
\hat \phi(\bk)^2}{2}\right)|0_I\rangle \,dk.
\end{eqnarray}
For the case of a scalar field, one has, for a mode $\bk$, $F_S=a^{-3}$,
$G_S=a^3\omega_k^2$ and the Ermakov-Pinney equation (\ref{Pinneyeq}) can be solved exactly with
\be\label{xscalar}
x_S=\sqrt{\frac{\pi H_\nu^{(1)}(z)H_{\nu^*}^{(2)}(z)}{2\Hds}}
\ee
where the $H_\nu^{(i)}(z)$ are the Hankel functions and
$\nu=\sqrt{\frac{9}{4}-\frac{m^2}{\Hds^2}}$. We note that the solution
(\ref{xscalar}) has been chosen to fulfil the requirement that $|0_I\rangle$
coincides with Bunch-Davies vacuum. One can finally
calculate $E_0^S(k)$ exactly and evaluate it in the $z>>1$ limit
\begin{eqnarray}\label{Hscalareta}
E_0^S(k)\stackrel{z\gg 1}{=}&&\!\!\!\!\!\frac{k}{2\,a}+\frac{(\Hds^2+m^2)\,a}{4\,k}\nonumber \\
&&\!\!\!\!\!+\frac{m^2(2\Hds^2-m^2)\,a^3}{16\,k^3}+o\left(\frac{1}{z^{3}}\right).
\end{eqnarray}
%%%%%%%%%%%%%%%%%%%%%%%%%%%%%%%%%%%
\subsubsection{Real Vector Field}
The Proca action for a minimally coupled, massive vector field
$A_\mu(\bx,t)\equiv (A_0,\vec A)$ is given by 
\be\label{Svector}
S\left[A_\mu\right]=\int d^4x \sqrt{-g} \left[ -\frac{1}{4} F_{\mu \nu} F^{\mu \nu} 
-\frac{M^2}{2}A^\mu A_\mu\right] 
\ee
where $F_{\mu\nu}\equiv\nabla_\mu A_\nu-\nabla_\nu A_\mu=\partial_\mu
A_\nu-\partial_\nu A_\mu$ (with $\nabla_\mu$ the covariant derivative).
We note that in the massive case the vector field satisfies the Lorentz gauge
condition $\nabla^\mu A_\mu=0$ and has three independent physical degrees of
freedom; in the massless case, instead, to the Lorentz condition one  should
add $A_0=0$, leading to  $\vec \nabla\cdot \vec A=0$ (radiation gauge) and
$A_\mu$ reduces to two independent physical components. The final expression
one  obtains for the action (\ref{Svector}), without any gauge fixing, is
\begin{eqnarray}\label{Svector2}
S\left[A_\mu\right]=&&\!\!\!\!\!\!\int d^4x \;\frac{a}{2}\left[\left(\dot{\vec
    A}\right)^2+ \left(\vec\nabla A_0\right)^2-2\dot{\vec A}\cdot\left(\vec
    \nabla A_0\right) \right.\nonumber\\
&&\!\!\!\!\!\!\left.-\left(\frac{\vec\nabla\wedge\vec A}{a}\right)^2+
M^2\left(a^2A_0^2-\vec A^2\right)\right].
\end{eqnarray}
For the massive case, the variation of the action with respect to $A_0$ leads
to the Gauss constraint:
\be\label{gauss}
\frac{1}{a^2}\left[\vec \nabla\cdot \dot{\vec A}-\nabla^2A_0\right]+M^2A_0=0.
\ee
If one considers the Fourier transform of the field components
\be\label{fourvector}
A_\mu=\frac{1}{\sqrt{V}}\sum_\bk\alpha_\mu(\bk){\rm e}^{i\bk\cdot\bx}
\ee
then the action can be expressed in terms of
$\alpha_\mu(\bk)\equiv(\alpha_0(\bk),\vec\alpha(\bk))$ as
\begin{eqnarray}\label{Skvector}
S=&&\!\!\!\!\!\!\sum_\bk\int dt\,\frac{a}{2}\left[\dot{\vec\alpha}(\bk)\cdot
\dot{\vec\alpha}(-\bk)-2i\,\alpha_0(\bk)\,\dot{\vec\alpha}(-\bk)\cdot \bk \right.\nonumber \\
&&\!\!\!\!\!\!-\frac{\bk\wedge\vec\alpha(\bk)\cdot\bk\wedge\vec\alpha(-\bk)}{a^2}+
a^2\omega_k^2\alpha_0(\bk)\alpha_0(-\bk)\nonumber\\
&&\!\!\!\!\!\!-\left.M^2\vec\alpha(\bk)\cdot\vec\alpha(-\bk)\phantom{\frac{A}{B}}
\!\!\!\!\!\!\right]
\end{eqnarray}
with $\omega_k^2\equiv M^2+\frac{\bk^2}{a^2}$ and the Gauss equation
(\ref{gauss}) for each mode is
\be\label{kgauss}
\alpha_0(\bk)=-i\frac{\bk}{a^2\omega_k^2}\cdot\dot{\vec\alpha}(\bk).
\ee
The action (\ref{Skvector}) can now be written in terms of the components
$\vec\alpha(\bk)$. If one decomposes $\vec\alpha(\bk)$ into the sum of a longitudinal vector
\be\label{alphaL}
\vec \alpha_L(\bk)\equiv(\vec\alpha(\bk)\cdot\hat k)\,
\hat k=\frac{1}{\sqrt{2}}\left[u_L(\bk)+i\,v_L(\bk)\right]\,\hat k
\ee
with $\hat k=\bk/\sqrt{\bk^2}$, plus a transverse vector 
\begin{eqnarray}\label{alphaT}
\vec\alpha_T(\bk)=&&\!\!\!\!\!\vec\alpha(\bk)-\vec\alpha_L(\bk)\nonumber \\
=&&\!\!\!\!\!\frac{1}{\sqrt{2}}\sum_{i=1,2}\left[u_T^{(i)}(\bk)+i\,v_T^{(i)}(\bk)\right]
\hat e_i(\bk)
\end{eqnarray}
where the $\hat e_i(\bk)$ form an orthonormal basis, $\hat e_i(\bk)\cdot \hat
e_j(\bk)=\delta_{ij}$, for the transverse space, $\hat e_i(\bk)\cdot\hat k=0$,
the action (\ref{Skvector}) can finally be written as 
\begin{eqnarray}
\label{Skvector2}
S=&&\!\!\!\!\!\!\!\!\!\!\sum_{\bk^+}\left[\phantom{\sum_{i=1,2}}\!\!\!\!\!\!\!\!\!\!\!\!
\left(S_L[u_L^{\phantom{(i)}}\!\!\!(\bk)]+S_L[v_L(\bk)]\right)\right.\nonumber \\
&+&\!\!\!\!\!\left.\sum_{i=1,2}\left(S_T[u_T^{(i)}(\bk)]+S_T[v_T^{(i)}(\bk)]\right)\right]
\end{eqnarray}
where
\be\label{Sklong}
S_L[\varphi(\bk)]\equiv M^2\int dt\,\frac{a}{2}\left(\frac{1}{\omega_k^2}
\dot\varphi(\bk)^2-\varphi(\bk)^2\right)
\ee
and
\be\label{Sktras}
S_T[\varphi(\bk)]\equiv\int dt\,\frac{a}{2}\left(\dot\varphi(\bk)^2-
\omega_k^2\varphi(\bk)^2\right).
\ee
We note that, owing to the overall $M^2$ factor in front of (\ref{Sklong}),
one recovers the correct physical limit for the massless case: in fact only
the transverse contributions remain when $M\rightarrow 0$. In the latter case,
the full Hamiltonian vacuum expectation value (v.e.v.), in the continuum limit (\ref{contlim}), is given by
\begin{eqnarray}\label{Htras}
&&\!\!\!\!\!\!\!\!\!\!\!\!\frac{\langle 0_I|\hat H_T|0_I\rangle}{V}
\stackrel{V\rightarrow\infty}{\longrightarrow}
\frac{1}{\pi^2}\int\limits_0^\infty k^2 E_0^T(k)dk\nonumber \\
&&\!\!\!\!\!\!\!\!\!\!\!\!=\frac{1}{\pi^2}\int\limits_0^\infty k^2\langle 0_I|
\left(\frac{\hat \pi_T(\bk)^2}{2a}+
\frac{a\,\omega_k^2\,\hat \phi_T(\bk)^2}{2}\right)|0_I\rangle\, dk
\end{eqnarray}
where $\hat \pi_T(\bk)\equiv a\dot{\hat\phi}_T(\bk)$, the factor 2 accounts
for the
contributions coming from different transverse polarizations, and the
$M\rightarrow 0$ limit should be taken. In the massive case one has also the longitudinal part
\begin{eqnarray}\label{Hlong}
&&\!\!\!\!\!\!\!\!\!\!\!\!\!\!\!\!\frac{\langle 0_I|\hat H_L|0_I\rangle}{V}
\stackrel{V\rightarrow\infty}{\longrightarrow}\frac{1}{2\pi^2}
\int\limits_0^\infty k^2 E_0^L(k)dk\nonumber \\
&&\!\!\!\!\!\!\!\!\!\!\!\!\!\!\!\!=\frac{1}{2\pi^2}\int\limits_0^\infty
k^2\langle 0_I| \left(\frac{\omega_k^2\hat \pi_L(\bk)^2}{2aM^2}+
\frac{aM^2\hat \phi_L(\bk)^2}{2}\right)|0_I\rangle dk
\end{eqnarray}
with $\hat \pi_L(\bk)\equiv\frac{aM^2}{\omega_k^2}\dot{\hat \phi}_L(\bk)$.\\
The Ermakov-Pinney equation (\ref{Pinneyeq}) can be solved exactly for the
transverse part since $F_T=a^{-1}$, $G_T=a\,\omega_k^2$, and
\be\label{xtras}
x_T=\sqrt{\frac{\pi H_\nu^{(1)}(z)H_{\nu^*}^{(2)}(z)}{2\Hds}}
\ee
with $\nu=\sqrt{\frac{1}{4}-\frac{M^2}{\Hds^2}}$ is the solution which is associated
with the Bunch-Davies vacuum.
The vacuum energy, in the large $z$ limit, is given by
\be\label{Htraseta}
E_0^T(k)\stackrel{z\gg1}{=}\frac{k}{2\,a}+\frac{M^2\,a}{4\,k}-\frac{M^4\,a^3}{16\,k^3}+
o\left(\frac{1}{z^{3}}\right).
\ee
We note that quadratic and logaritmic divergent terms disappear in the massless case.\\ 
The longitudinal field equations are obtained by setting
$F_L=\frac{\omega_k^2}{aM^2}$, $G_L=a\,M^2$ and cannot be solved exactly;
however one can estimate the vacuum energy of (\ref{Hlong}) in the adiabatic,
$z>>1$ limit by expanding the solution of Eq. (\ref{Pinneypert}) as a  power series in $z$
\begin{eqnarray}\label{xlong}
x_L\stackrel{z\gg1}{=}&&\!\!\!\!\!\!\frac{1}{\sqrt{\Hds z}}
\left[1+\frac{2 \Hds^2-M^2}{4\Hds^2z^2}\right.\nonumber \\
&&\!\!\!\!\!\!\left.+\frac{5\left(M^4-12M^2\Hds^2+4\Hds^4\right)}
{32\Hds^4z^4}+o\left(\frac{1}{z^{4}}\right)\right]
\end{eqnarray}
and then expanding (\ref{vevH}) to obtain
\begin{eqnarray}\label{Hlongeta}
E_0^L(k)\stackrel{z\gg1}{=}&&\!\!\!\!\!\!\frac{k}{2\,a}+\frac{(\Hds^2+M^2)\,a}{4\,k}\nonumber \\
&&\!\!\!\!\!\!-\frac{M^2(6\Hds^2+M^2)\,a^3}{16\,k^3}+o\left(\frac{1}{z^{3}}\right).
\end{eqnarray}
Interestingly enough, the longitudinal vacuum energy is slightlty different
from that of a scalar field because of the  curved space kinematics.
%%%%%%%%%%%%%%%%%%%%%%%%%%%%%%%%%%%%%%%%%
\subsection{Fermionic Field}
The action for a massive spin $\frac{1}{2}$ field $\Psi(\bx,t)$ is:
\begin{eqnarray}\label{Sfermion}
S[\Psi]=&&\!\!\!\!\!\! \int d^4x\sqrt{-g} \left\{ \frac{i}{2}
\left[ \phantom{\sum}\!\!\!\!\!\!\!\overline{\Psi} \tilde\gamma^\mu
(\nabla_\mu \Psi) - (\overline{\nabla_\mu\Psi} )\tilde\gamma^\mu \Psi \right] \right.\nonumber \\
&&\!\!\!\!\!\! \left.- \mu \overline{\Psi} \Psi \phantom{\frac{1}{2}}\!\!\!\!\!\right\}  
\end{eqnarray}
where $\nabla_\mu$ is the gauge covariant derivative for spin $1/2$ fields, 
$\tilde\gamma^{\mu}\equiv\du{e}{\mu}{a}(x)\gamma^a$ and $\du{e}{\mu}{a}(x)$ are the vierbein.
In FRW space-time, on Fourier expanding 
\be
\Psi (\bx,t)= \sqrt{\frac{a^3}{V}} \sum_{\bk} 
e^{i {\bk} \cdot {\bx}}\,\psi(\bk,t),
\ee
the action (\ref{Sfermion}) can be rewritten in a compact form\cite{FGV} as 
\begin{equation}
S=\sum_\bk\int dt\,\psi(\bk)^\dagger\left[\frac{i}{2}
\overleftrightarrow{\partial_t}-\hat M(\bk)\right]{\psi}(\bk)
%-\dot{\psi}(\bk)^\dagger\psi(\bk)\right)\right.\nonumber \\
%&&\left.-\psi(\bk)^\dagger \hat M(\bk)\psi(\bk)\right]
\end{equation}
where $\hat M(\bk)$ is the $4\times 4$ matrix
\be\label{fermmat}
\hat M(\bk) = \left( \begin{array}{cc}
\mu & \vec{\sigma} \cdot \frac{\bk}{a} \\
\\
\vec{\sigma} \cdot \frac{\bk}{a}  & -\mu 
\end{array} \right) \,.
\ee
and $\vec \sigma$ are the standard $2\times 2$ Pauli matrices.
The Euler-Lagrange equation obtained for a spinor $\psi(\bk)$ is
\be\label{diraceq}
i\dot\psi(\bk)=\hat M(\bk)\psi(\bk)
\ee
and the Hamiltonian for the system can be easily calculated in terms of $\psi(\bk)$ as
\be\label{Hspinor}
H_f\equiv \sum_\bk H_f(\bk)=\sum_\bk \psi(\bk)^\dagger\hat M(\bk)\psi(\bk).
\ee
The general solution of the equation of motion (\ref{diraceq}) can be written
as a superposition of the 4 independent solutions
\be
w_{1}^{(r)}(\bk,t) = \sqrt{\frac{\pi\,z\, N_f}{2}}\left(
\begin{array}{c}
i H_{\nu}^{(1)}(z)\\
\\
\frac{\vec \sigma \cdot \bk}{k} H_{\nu -1}^{(1)}(z)
\end{array} \right) \chi ^{(r)}
\label{uno}
\ee
with $r=1,2$
and
\be
w_{2}^{(r)}(\bk,t) = -\sqrt{\frac{\pi\,z\, N_f}{2}} \left(
\begin{array}{c}
\frac{\vec\sigma \cdot \bk}{k} H_{\nu^* -1}^{(2)}(z)\\
\\
i H_{\nu^*}^{(2)}(z)
\end{array} \right)\chi ^{(r)}
\label{due}
\ee
with $r=3,4$,
where $\nu=\frac{1}{2}-i\frac{\mu}{\Hds}$, $N_f={\rm
  exp}\left[\pi\,\mu/\Hds\right]$, $r=1,3 (2,4)$ correspond to spin up (down)
for the Pauli spinors $\chi ^{(r)}$ and the $w_i^{(r)}$ depend on time through $z$.
Note that the independent solutions (\ref{uno},\ref{due}) reduce to the usual
static solutions in the $z\rightarrow+\infty$ limit and they may then be
associated with the time-independent creation-annihilation operators
$\hat b_r(\bk)$, $\hat b_r(\bk)^\dagger$, $\hat d_r(\bk)$, $\hat d_r(\bk)^\dagger$. 
One defines the Bunch-Davies vacuum $|0_{BD}\rangle$ by
\be\label{BDvacuum}
\hat b_r(\bk)|0_{BD}\rangle=\hat d_r(\bk)|0_{BD}\rangle=0
\ee
with
\be\label{commfermion}
\left\{\hat b_r(\bk),\hat b_r(\bk)^\dagger\right\}=
\left\{\hat d_r(\bk),\hat d_r(\bk)^\dagger\right\}=1.
\ee
In terms of the above operators the Heisenberg field operator $\hat \psi
(\bk,t)$ can be written as
\begin{eqnarray}\label{psik}
\hat \psi(\bk,t)=&&\!\!\!\!\!\!\sum_{r=1,2}\left[w_{1}^{(r)}(\bk,t)\,
\hat b_r(\bk)\right.\nonumber\\
&&\!\!\!\!\!\!\left.+(-1)^{r+1} w_{2}^{(r)}(\bk,t)\,\hat d_r(-\bk)^\dagger\right]
\end{eqnarray}
and the vacuum expectation value of the Hamiltonian (\ref{Hspinor}) can be
evaluated in the continuum limit (\ref{contlim})
%\begin{widetext}
\begin{eqnarray}\label{vevHfermion}
&&\frac{\langle 0_{BD} |\hat H_F|0_{BD}\rangle}{V}
\stackrel{V\rightarrow\infty}{\longrightarrow}\frac{1}{\pi^2}
\int\limits_0^\infty k^2 E_0^F(k)dk\nonumber \\
&&=\frac{1}{2\pi^2}\int\limits_0^\infty k^2dk\,\langle 0_{BD}|\hat H_f(\bk)|0_{BD}\rangle
\end{eqnarray}
%\end{widetext}
by using (\ref{BDvacuum}-\ref{psik}).
We note that each $\hat H_f(\bk)$ contains  the contributions both of a
particle and an antiparticle with opposite spin and momentum; in terms of the
Hankel functions, one obtains the following exact expression:
\begin{eqnarray}\label{exactvevHferm}
\!\!\!\langle 0_{BD}|\hat
H_f(\bk)|0_{BD}\rangle=&&\!\!\!\!\!\!2\left[m\left(H_{\nu-1}^{(1)}
H_{\nu^*-1}^{(2)}-H_\nu^{(1)}H_{\nu^*}^{(2)}\right)\right.\nonumber\\
&&\!\!\!\!\!\!\!\left.+i\frac{k}{a}\left(H_{\nu-1}^{(1)}H_{\nu^*}^{(2)}-
H_\nu^{(1)}H_{\nu^*-1}^{(2)}\right)\right]\nonumber\\
&&\!\!\!\!\!\!\times\left(H_\nu^{(1)}H_{\nu^*}^{(2)}+H_{\nu-1}^{(1)}H_{\nu^*-1}^{(2)}\right)^{-1}
\end{eqnarray}
and, on counting a single particle/antiparticle degree of freedom, 
in the large $z$ limit, one finds
\be\label{vevFermion}
E_0^{F}(k)=-\frac{k}{2\,a}-\frac{\mu^2\,a}{4\,k}+
\frac{\mu^2(\Hds^2+\mu^2)\,a^3}{16\,k^3}+o\left(\frac{1}{z^{3}}\right).
\ee
%%%%%%%%%%%%%%%%%%%%%%%%%%%%%%%%%%%%%%%%%%%%%%%%%%%%%%%%%%%%%%%%%%%%%%%%%%%%%%%%%%%
\section{Sum rules and cancellation of ultraviolet divergences}
In this section we consider the general condition for the cancellation of
divergent vacuum contributions to the Einstein equations coming from different
free fields using the K\"allen-Lehmann spectral functions formalism.
Some specific models will then be analyzed in detail.
For the Minkowski case one can obtain some general results which will later be
generalized in a non trivial way to the de Sitter case.
%%%%%%%%%%%%%%%%%%%%%%%%%%%%%%%%%%%%%%
\subsection{Spectral function general formalism}
Starting from a semiclassical approach we consider the Friedmann equation for
the homogeneous scale factor $a(t)$ in (\ref{rwmetric})
\be\label{Eins}
a^3 \Hds^2=\frac{8\pi G}{3}\sum_j\frac{\langle 0_{BD}|\hat H_j|0_{BD}\rangle}{V}
\ee
where $\sum_j$ is the sum over all bosonic and fermionic field (representing
the particle content evolving on the space-time manifold) Hamiltonians
averaged with respect to the corresponding Bunch-Davies vacua.
The spectral function $\rho_A(x)$ can be introduced if we replace the above sum by
\begin{eqnarray}\label{sums}
\sum_j\rightarrow&&\!\!\!\!\!\frac{1}{2\pi^2}
\int\limits_0^\infty k^2dk\left[\int dm^2\rho_S(m^2)\right.\nonumber\\
&&\!\!\!\!\!\left.+\int dM^2\left(\rho_V(M^2)-v_0\delta(M^2)\right)\right.\nonumber\\
&&\!\!\!\!\!\left.+\int dM^2\,2\,\rho_V(M^2)\right.\nonumber\\
&&\!\!\!\!\!\left.+\int d\mu^2\,2\,\rho_F(\mu^2)\right]
\end{eqnarray}
where $\rho_A(x)$ have support for $0\le x<\infty$, $v_0\delta(M^2)$ is the
number density of massless vector particles, and Dirac spinors account for double the
contribution of Majorana and Weyl particles.
Moreover the second line of Eq. (\ref{sums}) counts the longitudinal
vector degrees of freedom while the third line gives the transverse contributions. 
On a de Sitter background, setting $m^2\equiv x$ with $\Lambda$ being the
ultraviolet cutoff ($ k\le a H \Lambda$) we adopt to regularize the divergent
integral $\int_0^\infty dk$\ , Eq. (\ref{Eins}) can be finally rewritten as
\begin{widetext}
\begin{eqnarray}\label{Eins2}
\!\!\!\!\!\!\!\!\frac{3\pi}{4G}=&&\!\!\!\!\!\Hds\int\limits_0^{\Lambda} z^2dz
\int dx\left[\rho_S(x)E_0^S(x,z)+\left(\rho_V(x)-v_0\delta(x)\right)
E_0^L(x,z)+2\,\rho_V(x)\,E_0^T(x,z)+2\,\rho_F(x)\,E_0^F(x,z)\right]\nonumber\\
=&&\int dx\left(\Lambda^4\frac{\Hds^2}{8}F_1(x)+\frac{\Lambda^2}{8}F_2(x)+
\frac{\log\Lambda}{16\Hds^2}F_3(x)+\mathcal{O}\left(\frac{1}{\Lambda^2}\right)\right)\, ,
\end{eqnarray}
\end{widetext}
where
\be\label{div4}
F_1(x)=\rho_S(x)+3\,\rho_V(x)-v_0\delta(x)-2\,\rho_F(x)\,,
\ee
\begin{eqnarray}\label{div2}
\!\!\!\!\!\!\!\!\!F_2(x)=\!\!\!\!\!&&\left(\Hds^2+x\right)\rho_S(x)+
\left(3\,x+\Hds^2\right)\,\rho_V(x)\nonumber\\
\nonumber\\
&&\!\!-v_0\left(x+\Hds^2\right)\delta(x)-2\,x\, \rho_F(x),
\end{eqnarray}
\begin{eqnarray}\label{divlog}
\!\!\!\!\!\!\!\!F_3(x)=\!\!\!\!\!&&x\left(2\Hds^2-x\right)\rho_S(x)-
x\left(6\Hds^2+3x\right)\,\rho_V(x)\nonumber\\
\nonumber\\
&&\!\!+v_0x\left(6\Hds^2+x\right)\delta(x)+2\,x\left(\Hds^2+x\right) \rho_F(x).
\end{eqnarray}
In order to obtain a finite contribution on the right hand side of
Eq. (\ref{Eins2}), the spectral functions describing the particle content of
the universe should be such as to make (\ref{div4}-\ref{divlog}) vanish simultaneously and
this occurs when
\begin{eqnarray}
\int dx F_1(x)=\!\!\!\!\!&&0\label{c1}\\
\nonumber\\
\int dx\,x\,F_1(x)=\!\!\!\!\!&&\Hds^2\int dx\left(\phantom{\frac{A}{B}}
\!\!\!\!\!\!\!v_0\delta(x)-\rho_S(x)-\rho_V(x)\right)\label{c2}\\
\nonumber\\
\int dx\,x^2F_1(x)=\!\!\!\!\!&&2\Hds^2\int dx\,x\left(\phantom{\frac{A}{B}}
\!\!\!\!\!\!\!\rho_S(x)-3\,\rho_V(x)+3\,v_0\delta(x)\right.\nonumber\\
&&\left.+\rho_F(x)\phantom{\frac{A}{B}}\!\!\!\!\!\!\!\right)\label{c3}.
\end{eqnarray}
In the $\Hds\rightarrow 0$ limit the conditions (\ref{c1}-\ref{c3}) reduce to
the very compact form 
\be
\int dx\,x^i\,F_1(x)=0,\quad i=0,1,2.
\label{cimink}
\ee
We note that the expression (\ref{Eins2}) is completely general and refers to
arbitrary mass distributions. In the following sections we shall analyze models
wherein the mass distributions are discrete and the $\rho_i(x)$ are just
superpositions of Dirac delta functions multiplied by integer coefficients,
$\rho_A=\sum_i n_i(A)\delta(x-x_i)$, and describe diverse particle
multiplets.

For $i=0$ Eq. (\ref{cimink}) gives the condition of the equality of numbers of
bosonic and fermionic degrees of freedom. For $i=1,2$ one has
\begin{eqnarray}
&&\sum m_s^2 +3 \sum m_V^2 = 2 \sum m_F^2\,
\nonumber\\
&&\sum m_s^4 +3 \sum m_V^4 = 2 \sum m_F^4\nonumber \, 
\end{eqnarray}
which can be regarded as hyperplane and hypersphere equations respectively
in term of squared mass variables, provided the bosonic or fermionic field
content is fixed.
%%%%%%%%%%%%%%%%%%%%%%%%%%%%%
\subsection{Hypersphere and hyperplane}
While the cancellation of the quartic divergences requires the equality of the
numbers of fermionic and bosonic degrees of freedom,
the conditions for the cancellation of quadratic and logarithmic divergences
are more involved. However, {\it for the Minkowski case}
these conditions can be represented as hyperplane and hypersphere equations respectively. 

Let us consider the hypersphere of dimensionality $n-1$ embedded into the
space of $n$ dimensions, given
by the equation
\begin{equation}
\sum_{i=1}^n y_i^2 = 1
\label{hypersphere}
\end{equation}
and the hyperplane given by the equation
\begin{equation}
\sum_{i=1}^n y_i = \lambda.
\label{hyperplane}
\end{equation}
We look for the points of intersection between the hypersphere and
hyperplane such that all the coordinates $y_i$ have non-negative values.
Then the minimum value of the parameter $\lambda$ is $\lambda = 1$ which
corresponds to the case when only one of coordinates is non-zero, while the
maximum value of $\lambda$ is $\lambda = \sqrt{n}$, when all
the coordinates are equal ($y_i = \frac{1}{\sqrt{n}}$, for $\forall i$). Thus
\begin{equation}
1 \leq \lambda \leq \sqrt{n}.
\label{range}
\end{equation}

The intersection between the hypersphere (\ref{hypersphere}) and the
hyperplane (\ref{hyperplane}) is a $(n-2)$-dimensional
hypersphere, whose center has equal coordinates $y_i$ for all $i$ and radius
$\sqrt{1-\lambda^2/n}$~. 
All the points of the hypersphere of intersection $S^{(n-2)}$ can be
represented by the vector
\begin{equation}
\frac{\lambda}{n} \vec{v} + \sqrt{1 -\frac{\lambda^2}{n}} \vec{u},
\label{vector}
\end{equation}
 where
\begin{eqnarray}
&&\vec{v} \equiv (1,\cdots,1),\\
\label{vectorv}
\nonumber\\
&& \vec{u} \cdot \vec{u} = 1, \; \vec{u} \cdot \vec{v} = 0.
\label{vectoru1}
\end{eqnarray}
The coordinates of the points, belonging to $S^{(n-2)}$ are
\begin{equation}
y_i = \frac{\lambda}{n} + \sqrt{1 -\frac{\lambda^2}{n}} u_i.
\label{coordinate}
\end{equation}

Let us find the minimum and maximum possible values of $u_i$. Without losing
generality we will focus on $i = 1$.
We shall look for a solution in the form of a vector
\begin{equation}
\vec{u} = (\alpha,\vec{\beta}),
\label{vectormax}
\end{equation}
where $\beta$ is a $(n-1)$-dimensional vector.
The conditions (\ref{vectoru1}) give 
\begin{equation}
\alpha^2 + \vec{\beta} \cdot \vec{\beta} = 1
\label{vectormax1}
\end{equation}
and
\begin{equation}
\alpha + \vec{\beta} \cdot \vec{I} = 0,
\label{vectormax2}
\end{equation}
where $I$ is a $(n-1)$-dimensional vector with all the coordinates
equal to $1$. 
Then
\begin{equation}
\vec{\beta} \cdot \vec{\beta} \, \left( (n-1)\cos^2\chi + 1\right) = 1.
\label{vectormax6}
\end{equation}
where $\chi$ is an angle between the vectors $\vec{\beta}$ and
$\vec{I}$. 
Apparently, $\alpha^2$ has maximum value 
when the angle
$\cos \chi = \pm 1$. Then
\begin{equation}
\alpha = \pm \sqrt{\frac{n-1}{n}}, \label{alpha}
\end{equation}
\begin{equation}
\vec{\beta} = \mp \frac{1}{\sqrt{n(n-1)}} \vec{I}.
\label{beta}
\end{equation}
Substituting the maximum and minimum values of $\alpha$ from Eq.
(\ref{alpha}) into Eq. (\ref{coordinate}) one can find the maximum
and minimum values of one of the coordinates $y_i$:
\begin{equation}
y_{i\ max} = \frac{\lambda}{n} +
\sqrt{1-\frac{\lambda^2}{n}}\sqrt{\frac{n-1}{n}},
\label{maximal}
\end{equation}
\begin{equation}
y_{i\ min} = \frac{\lambda}{n} -
\sqrt{1-\frac{\lambda^2}{n}}\sqrt{\frac{n-1}{n}}. \label{minimal}
\end{equation}
We note that when one of the coordinates $y_i$ takes an extremum value,
all the other coordinates have equal values (see Eq. (\ref{beta})).
Now, on using Eq. (\ref{minimal}) we can get the condition for the
positivity of all the coordinates $y_i$ of the points lying on the
$(n-2)$ - dimensional sphere. Requiring $y_{i\ min} \geq 0$ we have
\begin{equation}
\lambda \geq \sqrt{n-1}.
\label{positive}
\end{equation}
Curiously, when one of coordinates acquires the maximum
value (\ref{maximal}) all the other coordinates have value
\begin{equation}
y_j = \frac{\lambda}{n} -
\sqrt{1-\frac{\lambda^2}{n}}\frac{1}{\sqrt{n(n-1)}},
\label{positive1}
\end{equation}
which is non-negative for $\lambda \geq 1$ and hence is always
non-negative for our range of $\lambda$.

Let us now consider the case $\lambda < \sqrt{n-1}$. In this case
some of the points of $S^{(n-2)}$ have negative
values of coordinates and should be excluded from consideration.
These patches of the hypersphere have the form of $n$ spherical caps,
whose pole points are the points where one of $n$ coordinates has
its minimum and, hence, for $\lambda < \sqrt{n-1}$ negative value.
The angle characterizing the size of
these spherical caps (i.e. the angle between the unit vector pointing to the
pole and the vector pointing to the intesection of
$S^{(n-2)}$ with the hyperplane $y_i = 0$) is  
\begin{equation}
\theta = \arccos \frac{\lambda}{\sqrt{(n-\lambda^2)(n-1)}}.
\label{angle1}
\end{equation}
One can find also the angle between two vectors pointing to two different ``poles'':
\begin{equation}
\chi = \arccos \left(-\frac{1}{n-1}\right).
\label{angle2}
\end{equation}
In the case for which $2\theta > \chi$ two spherical caps can intersect each
other. A simple calculation shows that it occurs if 
$\lambda < \sqrt{n-2}$. 
Let us note that this does not mean that acceptable solutions do not exist.
Indeed, as we explained above, when one of the coordinates $y_i$ has the maximum value (\ref{maximal}) all the other
coordinates have non-negative values (\ref{positive1}): for the case $\lambda > 1$ these values are positive.
%%%%%%%%%%%%%%%%%%%%%%%%%%%%%%%%%%%%%%
\subsection{Some simple models}
In our case the parameter $\lambda$ is nothing more than
\begin{equation}
\lambda = \frac{4\sum_{i=1}^{D} m_{iD}^2 + 2\sum_{i=1}^{M}
  m_{iM}^2}{\sqrt{4\sum_{i=1}^{D} m_{iD}^4 + 2\sum_{i=1}^{M} m_{iM}^4}},
\label{fermions}  
\end{equation}
where $m_{iD}$ and $m_{iM}$ are the masses of Dirac and Majorana spinors
respectively, while $D$ and $M$ are the numbers of these 
spinors. The number $n$ is the total number of fermionic degrees of freedom 
\begin{equation}
n = 4D + 2M + 2W,
\label{fermnions1}
\end{equation}
where $W$ is the number of Weyl spinors.
The number of bosonic degrees of freedom should be equal to the number of fermionic 
degrees of freedom $n$ and the boson masses (squared) are given by
$y_i \sqrt{4\sum_{i=1}^{D} m_{iD}^4 + 2\sum_{i=1}^{M} m_{iM}^4}$.
It is easy to see that the value of $\lambda$ cannot
take the minimum value $\lambda = 1$ (cf. Eq. (\ref{range})). 

Let us consider the model with two degrees of freedom ($n =2$). In this case
we have a Majorana spinor and the value of the parameter 
$\lambda$ is $\lambda = \sqrt{2}$ and the circumference and straight line have
the only point of intersection, when $y_1 = y_2$. Obviously this situation
corresponds to to two scalars with the same mass, which coincides with that of the 
Majorana spinor. Therefore in this case the conditions to have the required cancellations
are so severe that the only possible solution is the one which can be obtained by
imposing supersymmetry for bosons and fermions. We shall see that
for larger systems these constraints are not so stringent.
If instead of Majorana one has a Weyl spinor,
the bosonic part of the spectrum will be represented by 
two massless scalar fields or by one electromagnetic field.  

The model with the four degrees of freedom represented by one Dirac spinor or
by two Majorana spinors with identical 
masses ($n = 4, \lambda = 2$) also
has only a trivial solution: four scalar fields with the same mass or one
scalar field and one massive vector field 
whose masses coincide. 
The case with two Majorana spinors with different masses is more interesting.
In this case $n = 4$, and 
\begin{equation}
\lambda = \sqrt{2} \frac{m_1^2 + m_2^2}{\sqrt{m_1^4+m_2^4}}\,.
\label{lambda-m}
\end{equation}

Now on using the results of the preceding section it is easy to see that the prohibited caps 
do not exist if $\lambda \geq \sqrt{3}$ (see Eq. (\ref{positive})).
Substituting this condition into Eq. (\ref{lambda-m}) one 
re-expresses this condition in terms of the relation between two fermion masses:
\begin{equation}
2 + \sqrt{3} \geq m_1^2/m_2^2 \geq 2 - \sqrt{3}.
\label{proh} 
\end{equation}
If this condition is violated the prohibited caps exist, but they do not intersect each other. 

In the following we shall investigate some specific models by directly studying the
conditions (\ref{c2}) and (\ref{c3}) and using a slightly different notation:
$x_i$ and $y_i$ will now denote the square of the boson and fermion masses
respectively.

Let us consider a particular case with two fields in the bosonic sector:
one scalar and one vector massive fields are present.
Now, the sum rules lead to
\begin{equation}
3 x_1 + x_2 = 2 y_1 + 2 y_2,
\label{balance}
\end{equation}
\begin{equation}
3 x_1^2 + x_2^2 = 2 y_1^2 + 2 y_2^2,
\label{balance1}
\end{equation}
where $x_1$ and $x_2$ are the vector boson and scalar masses squared, while $y_1$
and $y_2$ are the Majorana spinor masses squared 
($y_2 \geq y_1$). On solving Eqs. (\ref{balance}) and (\ref{balance1}) we get 
\begin{equation}
x_1 = \frac{y_1 + y_2}{2} \pm \frac{\sqrt{3}}{6}(y_2 - y_1),
\label{balance2} 
\end{equation}
\begin{equation}
x_2 = \frac{y_1 + y_2}{2} \mp \frac{\sqrt{3}}{2}(y_2 - y_1).
\label{balance3} 
\end{equation}
Let us note that the first solution for $x_2$ is negative if $\frac{y_1}{y_2} =
\frac{m_1^2}{m_2^2} \leq 2 - \sqrt{3}$. 
Obviously this condition coincides with the condition (\ref{proh}). This is
quite natural, since the case with one scalar and one 
vector massive fields corresponds to the vector $\vec{u}$, considered in the
preceding section, having an extremum value component while all the
others have equal values. Thus, when the condition (\ref{positive}) is satisfied there are two 
solutions for the system of equations (\ref{balance}) and (\ref{balance1}),
corresponding to the scalar field having maximum and minimum values,
while if this condition is not valid only one solution survives.   

Let us now consider the model with 6 degrees of freedom, wherein the fermion
sector is represented by one Dirac and one Majorana spinors. To begin with,
consider the bosonic sector containing two massive vector fields. The sum rules
in this case lead to
\begin{equation}
3 x_1 + 3 x_2 = 4 y_1 + 2 y_2,
\label{balance4}
\end{equation}
\begin{equation}
3 x_1^2 + 3 x_2^2 = 4 y_1^2 + 2 y_2^2.
\label{balance5}
\end{equation}
where $y_1$ and $y_2$ represent the Dirac and Majorana fermion masses squared. 
The solution of equations (\ref{balance4}) and (\ref{balance5}) is 
\begin{equation}
x_{1,2} = \frac{2 y_1 + y_2 \pm \sqrt{2} |y_1 - y_2|}{3}.
\label{balance6} 
\end{equation}
This solution is positive for both $x_1$ and $x_2$ provided the following condition is satisfied:
\begin{equation}
\frac{m_2^2}{m_1^2} = \frac{y_2}{y_1} \leq 4 + 3\sqrt{2}.
\label{Dir-Maj}
\end{equation}
 
Let us now consider the bosonic sector which contains a massive vector field,
an electromagentic field and a scalar field. In this case the sum rules are 
\begin{equation}
3 x_1 + x_2 = 4 y_1 + 2 y_2,
\label{balance7}
\end{equation}
\begin{equation}
3 x_1^2 + x_2^2 = 4 y_1^2 + 2 y_2^2.
\label{balance8}
\end{equation}
The existence of positive solutions for $x_1$ and $x_2$ depends on the
relation between fermion masses. Namely, if 
\begin{equation}
\frac{y_1}{y_2} \leq \frac{3\sqrt{2}-4}{2}
\label{Dir-Maj1}
\end{equation}
there is one solution 
\begin{equation}
x_1  = y_1 + \frac{y_2}{2} - \sqrt{\frac{y_2^2}{12} - \frac{y_1y_2}{3}},
\label{balance9}
\end{equation}
\begin{equation}
x_2  = y_1 + \frac{y_2}{2} + 3\sqrt{\frac{y_2^2}{12} - \frac{y_1y_2}{3}}.
\label{balance10}
\end{equation}  
If 
\begin{equation}
\frac{3\sqrt{2}-4}{2} \leq \frac{y_1}{y_2} \leq \frac{1}{4}
\label{Dir-Maj2}
\end{equation}
then two solutions exist. One of them is the solution (\ref{balance9}), (\ref{balance10}) and 
the second solution is 
\begin{equation}
x_1  = y_1 + \frac{y_2}{2} + \sqrt{\frac{y_2^2}{12} - \frac{y_1y_2}{3}},
\label{balance11}
\end{equation}
\begin{equation}
x_2  = y_1 + \frac{y_2}{2} - 3\sqrt{\frac{y_2^2}{12} - \frac{y_1y_2}{3}}.
\label{balance12}
\end{equation} 
Finally, if $\frac{y_1}{y_2} > \frac{1}{4}$ solutions do not exist.
%%%%%%%%%%%%%%%%%%%%%%%%%%%%%%%%%%%%%%%%%%%%%%%
\subsection{Simple models in the presence of the cosmological constant}
For the case of the cosmological constant being different from zero, the relations
between masses of the fields, leading to the cancellation 
of divergences in the vacuum energy become more involved. 

Consider the case of two degrees of freedom. In this case the fermionic sector
is represented by one Majorana spinor with mass squared denoted by $y$,
while the bosonic sector is represented by two scalar fields with masses squared 
denoted by $x_1$ and $x_2$  The sum rules are 
\begin{equation}
x_1 + x_2 = 2 y - 2 \Hds^2,
\label{cosm}
\end{equation}
\begin{equation}
x_1^2 + x_2^2 - 2 \Hds^2 (x_1 + x_2) = 2 y (y + \Hds^2).
\label{cosm1}  
\end{equation}
Non-negative solutions of this system of equations are possible provided the
Majorana fermion mass satisfies the constraint 
\begin{equation}
y \geq  \frac{7 + \sqrt{33}}{2} \Hds^2.
\label{restrict}
\end{equation}
In this case the masses of the scalar fields are 
\begin{equation}
x_{1,2} = y - \Hds^2 \pm \sqrt{5 y \Hds^2 - 3 \Hds^4}.
\label{cosm2}
\end{equation}
For the case of 4 degrees of freedom, one can have as a non-trivial solution
for the fermionic sector a Dirac fermion. If the bosonic sector
is represented by an electromagnetic field and two scalar fields, the
sum rules are: 
\begin{equation}
x_1 + x_2 = 4 y - 2 \Hds^2,
\label{cosm3}
\end{equation}
\begin{equation}
x_1^2 + x_2 ^2 - 2 \Hds^2 (x_1 + x_2) = 4 y (y + \Hds^2)\,.
\label{cosm4}
\end{equation}
The system of equations (\ref{cosm3}) and (\ref{cosm4}) has non trivial
non-negative solutions and the expected condition $x_1=x_2=y=0$ is recovered as
$\Hds$ tends to zero. In the presence of non zero $\Hds$ if 
\begin{equation}
2 \Hds^2 \le y \le \frac{1}{2}\left(5+\sqrt{19}\right) \Hds^2 
\label{restrict1}
\end{equation}
one can easily obtain
\begin{equation}
x_{1,2} = 2 y - \Hds^2 \pm \sqrt{10 y \Hds^2 - 2 y^2 - 3 \Hds^4}
\label{cosm5}
\end{equation}
where the lower bound in \eqref{restrict1} is required for $x_{1,2}$ to be
positive while the upper bound is needed for the reality of the solutions.
 
The last case which we treat here is the bosonic sector represented by a
scalar and a vector massive fields. The sum rules take the following form;
\begin{equation}
3 x_1 + x_2 + 2 \Hds^2 = 4 y,
\label{cosm6}
\end{equation}
\begin{equation}
3 x_1^2 + x_2^2 + 2 \Hds^2 (3 x_1 - x_2) = 4 y (y + \Hds^2).
\label{cosm7}
\end{equation}
If the condition $y > 2\Hds^2$ is satisfied there are two non-negative solutions:
\begin{equation}
x_1 = y - \Hds^2 + \sqrt{\frac{\Hds^2(\Hds^2 + y)}{3}},
\label{cosm8}
\end{equation}
\begin{equation}
x_2 = y + \Hds^2 - 3 \sqrt{\frac{\Hds^2(\Hds^2 + y)}{3}},
\label{cosm9}
\end{equation}
and 
\begin{equation}
x_1 = y - \Hds^2 - \sqrt{\frac{\Hds^2(\Hds^2 + y)}{3}},
\label{cosm10}
\end{equation}
\begin{equation}
x_2 = y + \Hds^2 + 3 \sqrt{\frac{\Hds^2(\Hds^2 + y)}{3}}.
\label{cosm11}
\end{equation} 
In the limiting case $y = 2 \Hds^2$ only the solution (\ref{cosm8}) and
(\ref{cosm9}) is valid because in formula (\ref{cosm10}) the mass of the
massive vector boson vanishes and this breaks the equality of the boson and
fermion degrees of freedom.

As a consequence of our approach it appears that the particle mass spectrum
and the Hubble constant of the De Sitter space are related.
Of course one may attempt to introduce some dynamics and an evolving mass spectrum,
which however we expect to still be connected to the Hubble parameter.
We plan to return to this in the future.
%%%%%%%%%%%%%%%%%%%%%%%%%%%%%%%%%%%%%
\section{Conclusions}
In this paper we considered the problem of vacuum energy in quantum
field theory and cosmology. This is often associated with the so
called cosmological constant problem. Indeed, the contribution of quantum
vacuum fluctuations to the  energy-momentum tensor behaves as a cosmological
constant \cite{Zeldovich:1967gd}. However, the cosmological constant, 
in principle, can also be of non-quantum origin and be present in the theory as
one of fundamental constants. The real vacuum energy problem consists in
the absence of a well established and justified procedure for the
renormalization of ultraviolet divergences in the energy-momentum tensor
analogous to the theory of renormalization in standard quantum field theory in a 
flat (absence of gravity) space-time. Indeed, in quantum field theory
the ultraviolet divergences are absorbed in the renormalization of a finite
number of measurable constants, or, in other words, all the observable quantities become finite 
due to the introduction of infinite counterterms into the bare Langrangian
(see e.g. \cite{Bog-Shir,Weinbergbook}). As far the very strong
ultraviolet divergences in the energy-momentum tensor are concerned they are set equal to
zero by the normal or Wick quantization of quantum fields. The last step
could be justified by the fact that one always measures the differences
between energy levels and not the absolute values of energy.
However, this justification fails in the presence of gravity, because of the
very structure of the Einstein equations, which connects the curvature of
spacetime to the energy-momentum tensor. Thus, one should also take into account the
contribution of the vacuum expectation value of the energy-momentum tensor
on the right-hand side of Einstein equations. Indeed one may study models
wherein vacuum oscillations drive some stage of the expansion of
patches of the universe can be studied (see \cite{FVV}).

It is well known that the naive calculation of the vacuum energy of quantum
fluctuations using a cutoff on the Planck scale gives a huge value for
it. One has two possible ways out of this situation. The first
one is the construction of a consistent renormalization theory for the
ultraviolet divergences of the energy-momentum tensor, which is a very
difficult task, since we do not have such relatively simple tools as
the renormalization of some known physical quantities as, for example, in quantum 
electrodynamics (see, however, \cite{Bar-Kam}, where an attempt to fix
the renormalization of an effective cosmological constant based on some
self-consistency conditions was undertaken). On the other hand one may require
the exact cancellation of ultraviolet divergences in the energy-momentum
tensor. This is the approach attempted in the present paper
(a good review of analogous approaches is given in \cite{visser},
further we also wish to mention a related flat space approach in the context
of induced gravity \cite{frolov}).
Our idea can be simply formulated as follows: if you do not know what to do
with these ultraviolet divergences, just try to eliminate them by introducing the
condition of their cancellation as a quantum consistency condition of the theory. 
Indeed, the cancellation conditions on the spectral functions mantioned in the
Introduction and studied in the third section of this paper is one of the oldest
examples of such approach. Other well-known examples of quantum
self-consistency are connected with conditions for the cancellation of quantum
anomalies. A classical example is the mechanism suggested by Glashow, Illioloulos and Maiani 
\cite{GIM}, who introduced the fourth quark into the theory of
electroweak interactions to suppress the chiral anomaly.
Later this quark was discovered experimentally and was called
``charmed''. Another spectacular example is the appearence    
of critical dimensions in string and superstring theories
\cite{string}. The theory of a bosonic string can be formulated 
consistently in a spacetime of 26 dimensions, while the critical
dimensionality for superstrings is equal to 10. 
Analogous relations between the dimensionality of spacetime and its matter
content arise also in the application of the 
BFV-BRST quantization mechanism to quantum cosmology \cite{Kam-Lyakh}. Yet another
restriction on the particle spectrum of a theory 
based on the requirement of the normalizability of the wave function of the
universe was obtained in \cite{Bar-Kam-norm}. 
   
Thus, our work, which uses the condition of cancellation of ultraviolet
divergences in the vacuum energy to arrive to some special conditions on the
spectrum of the fundamental theory, is in kinship with a rather fruitful and ramified stream 
of the development of modern quantum theory. Let us now mention what are
the similarities and the differences between our approach and the supersymmetry one.
In the case of exactly supersymmetric models in flat spacetime there is an exact 
cancellation of the vacuum energy and not only of its divergent part. This
cancellation occurs because the fermion and boson 
contributions to the vacuum energy enter with opposite signs.
However, exact supersymmetry cannot be implemented for 
the construction of a realistic theory of elementary particles. Moreover
there are serious difficulties arising in the formulation of supersymmetry in
curved background spacetimes. Here we share one common feature with supersymmetry:
the equal number of fermionic and bosonic degrees of freedom, which is
indispensable for the cancellation of quartic ultraviolet divergences. The
general conditions for the cancellation of quadratic and logarithmic divergences
are much more flexible than the requiring of exact supersymmetry.
On analysing some simple examples we have seen that there are many opportunities 
of satisfying these conditions. At the same time we do not require the
cancellation of the finite part of the vacuum energy. It could be different from
zero and be responsible for the observable cosmological constant. 

We note that the sum rule constraints may generally allow for theories whose
particle spectrum fit the content of the stardard model in the low mass
sector.
It is widely believed that some kind of extension of the
standard model is needed and many of them, such as the supersymmetric ones, have indeed
been considered. Therefore our approach may be helpful to encode the minimum possible
number of constraints on the spectrum of the theory necessary for the
cancellation of the divergences of the vacuum energy, even in the presence of
an effective cosmological constant. 
So, one may think of some kind of grand unification theory, where sum rules
can be realized. Perhaps, such a theory should contain not only additional
supermassive gauge bosons but also some additional families of fermions. 
A straightforward extension of our analysis can be done for the case of extra
spatial dimensions, one has simply to take into account the dimension
dependent number of physical boson and fermion degrees of freedom and write
the general identity necessary to saturate the corresponding spectral function sum rules.  
We believe that all these topics deserve further investigation, not to mention
the introduction of higher spins.\\
{\bf Acknoledgement}\\
We wish to thank G. E. Volovik for useful correspondence.
 %%%%%%%%%%%%%%%%%%%%%%%%%%

\end{document}